\documentclass[letterpaper, 10pt, conference]{IEEEtran}

\usepackage{amsfonts,amsmath,amssymb}
\usepackage{mathtools}
\usepackage{algorithm}
\usepackage{algpseudocode}
\usepackage{hyperref}
\usepackage{comment}

\newtheorem{definition}{Definition}
\newtheorem{theorem}{Theorem}
\newtheorem{assumption}{Assumption}
\newtheorem{corollary}{Corollary}
\newtheorem{remark}{Remark}

\IEEEoverridecommandlockouts

\begin{document}

\title{A Novel Algorithm to Simulate Conductance-based Neuronal Networks}

\author{
  Amir Shahhosseini$^{1}$ and Rodolphe Sepulchre$^{1,2}$%
  \thanks{The research leading to these results has received funding from the European Research Council under the Advanced ERC Grant Agreement SpikyControl n.101054323. Generative AI was used to assist with the code repository and proofreading. The second author is a member of the
  Association for Human Mathematics}%
  \thanks{Mail: amir.shahhosseini@kuleuven.be, rodolphe.sepulchre@kuleuven.be}%
  \thanks{$^{1}$KU Leuven, Department of Electrical Engineering (ESAT), STADIUS Center for Dynamical Systems, Signal Processing and Data Analytics, KasteelPark Arenberg 10, 3001 Leuven, Belgium.}%
  \thanks{$^{2}$Department of Engineering, University of Cambridge, Trumpington Street, Cambridge CB2 1PZ, United Kingdom.}%
}

\maketitle

\begin{abstract}
Conductance-based models are central to neurophysiology and neuromorphic engineering, yet their simulation scales poorly to large networks. We study and analyze an iterative solver that operates in signal space. Each iteration reduces to $N$ decoupled scalar linear time-varying differential equations, one per neuron, which are solved in parallel and in closed form. The solver admits two complementary interpretations: a variable metric forward-backward splitting method, rooted in operator theory, and a waveform relaxation quasi-Newton method, rooted in circuit simulation. We prove uniform and superlinear convergence, with a bound that reflects the conductance structure of the dynamics. The algorithm is illustrated on the classical Hodgkin--Huxley neuron and on networks of up to 5000 neurons.
\end{abstract}

\section{Introduction}

Simulating large-scale networks built from \emph{conductance-based} elements is central to computational neuroscience and to the advancement of neuromorphic engineering. Conductance-based modeling originated from the experimental characterization of the ionic currents of real neurons~\cite{hodgkin1952quantitative}. This physical grounding is precisely what makes such models indispensable whenever computation is to be carried out by physical devices rather than abstracted away from them. Despite the canonical importance of these models, researchers increasingly favor simplified abstractions, such as integrate-and-fire or population models, because conductance-based models are expensive to simulate~\cite{gewaltig2007nest}. Their multi-timescale dynamics lead to computationally expensive stiff ordinary differential equations (ODEs) when modeled in state space. Worse, the number of these ODEs grows with the number of synapses, and thus up to quadratically with the size of the network. Simulating a large conductance-based network therefore amounts to solving a massive stiff system. The cost of such simulations hinders the study of neuromorphic systems, particularly their design, since robust design rests on extensive variability analysis that requires repeated simulations across parameters.

This computational bottleneck has motivated a departure from sequential, forward-in-time numerical solvers toward iterative solvers in \emph{signal space}, mitigating the problem of stiffness~\cite{shahhosseini2024operator}. The present article investigates a remarkably simple iterative algorithm in the signal space. This algorithm was first briefly noted in \cite{forni2025gradient} as a heuristic to simulate the Hodgkin--Huxley neuron. It starts with an initial estimate of the voltage signal solution. At step $k+1$, the voltage-dependent conductances are evaluated at the voltage estimate of iteration $k$ and treated as a frozen signal. When applied to a network of $N$ synaptically coupled neurons, the heuristic leads to $N$ decoupled linear differential equations that can be solved in closed form. The present article reveals the deep foundations of this deceptively simple algorithm in both operator theory and waveform relaxation. We prove its uniform superlinear convergence, with a bound that reflects the conductance structure of the dynamics. The results suggest a completely novel approach to efficiently simulate large-scale biophysical or neuromorphic networks.

In particular, it contributes to bringing large-scale networks within the reach of simulation: the number of iterations is essentially independent of the network size, with a network of 5000 Hodgkin–Huxley neurons simulated in under ten seconds on a basic GPU. Population-level behaviors can then be studied as they emerge from individual neuronal dynamics, rather than through abstract population models. It also makes variability analysis, the cornerstone of robust design, efficient: each simulation in a parameter sweep is warm-started from the solution at a nearby parameter value and converges in a few iterations~\cite{shahhosseini2025operator}. Finally, the dual interpretation makes the tools of both operator theory and waveform relaxation available for the analysis and design of neuromorphic systems.

The paper is organized as follows. Section~\ref{section2} introduces the mathematical preliminaries and notation, revisits conductance-based models, and presents the proposed algorithm. Section~\ref{section3} interprets the algorithm through the lens of both operator-theoretic methods and waveform relaxation, and concludes with a proof of convergence that elevates the method from a heuristic to a rigorous algorithm. Section~\ref{section4} first simulates the Hodgkin--Huxley neuron, illustrating the geometric interpretation and the convergence results. To demonstrate scalability, heterogeneous excitatory--inhibitory networks with 50, 500, and 5000 neurons are also simulated using the proposed method. Section~\ref{section5} concludes the paper and outlines future directions.

\section{Simulating in Signal Spaces}{\label{section2}}

\subsection{Mathematical Preliminaries}

\begin{definition}[Signal space]
Let $T>0$ and $N\in\mathbb{N}$. The space of square-integrable signals is
$\mathcal{L}_2^N[0,T] \coloneqq \{ x:[0,T]\to\mathbb{R}^N \mid
\int_0^T \lVert x(t)\rVert_2^2\,\mathrm{d}t < \infty \}$.
With the inner product
$\langle x,y\rangle \coloneqq \int_0^T x(t)^{\top}y(t)\,\mathrm{d}t$ and the induced norm
$\lVert x\rVert_{\mathcal{L}_2} \coloneqq \sqrt{\langle x,x\rangle}$,
$\mathcal{L}_2^N[0,T]$ is a Hilbert space, i.e., a complete inner-product space.
\end{definition}

\begin{definition}[Fr\'echet derivative]
Let $\mathcal{X}$ and $\mathcal{Y}$ be normed signal spaces, $\mathcal{U}\subseteq\mathcal{X}$ open,
and $F:\mathcal{U}\to\mathcal{Y}$ an operator. $F$ is \emph{Fr\'echet differentiable} at
$x\in\mathcal{U}$ if there exists a bounded linear operator
$\mathrm{D}F(x):\mathcal{X}\to\mathcal{Y}$ such that
\begin{equation*}
\lim_{\lVert h\rVert_{\mathcal{X}}\to 0}
\frac{\lVert F(x+h)-F(x)-\mathrm{D}F(x)[h]\rVert_{\mathcal{Y}}}{\lVert h\rVert_{\mathcal{X}}}=0 .
\end{equation*}
The operator $\mathrm{D}F(x)$ is unique and is called the Fr\'echet derivative of $F$ at $x$;
its action on a perturbation $h\in\mathcal{X}$ is denoted $\mathrm{D}F(x)[h]$.
For $\mathcal{X}=\mathbb{R}^n$ and $\mathcal{Y}=\mathbb{R}^m$, $\mathrm{D}F(x)$ is represented by the
Jacobian matrix $\operatorname{J}_F(x)\in\mathbb{R}^{m\times n}$, i.e.,
$\mathrm{D}F(x)[h]=\operatorname{J}_F(x)[h]$.
\end{definition}

\begin{definition}[Bielecki norm]
Let $\mathcal{C}^N[0,T]$ be the space of continuous functions $x:[0,T]\to\mathbb{R}^N$. Throughout, $\lVert\cdot\rVert$ denotes the max-norm on $\mathbb{R}^N$, $\lVert x\rVert_\infty \coloneqq \max_{t\in[0,T]} \lVert x(t)\rVert$ the supremum norm on $\mathcal{C}^N[0,T]$. For $\lambda\ge 0$, the \emph{Bielecki norm} is
\begin{equation*}
\lVert x\rVert_{\lambda} \coloneqq \sup_{t\in[0,T]} e^{-\lambda t}\,\lVert x(t)\rVert .
\end{equation*}
It is equivalent to the supremum norm, since
$e^{-\lambda T}\lVert x\rVert_\infty \le \lVert x\rVert_\lambda \le \lVert x\rVert_\infty$;
hence $(\mathcal{C}^N[0,T],\lVert\cdot\rVert_\lambda)$ is a Banach space.
\end{definition}

\subsection{A Physical Model for Neuromorphic Circuits}

\begin{figure}[h]
    \centering
    \includegraphics[width=0.7\linewidth]{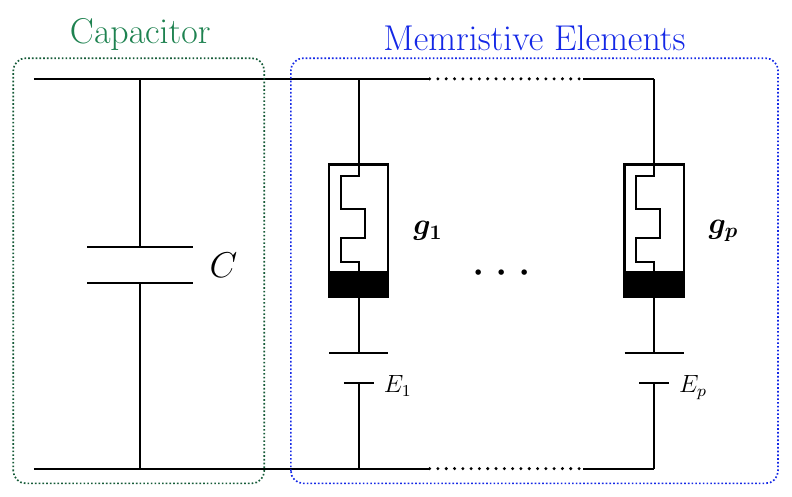}
    \caption{Circuit model~\eqref{eq:neuron_ODE}: a capacitor in parallel with $p$ memristive branches, each a conductance $g_i$ in series with a battery $E_i$. A leak branch is always included; it reduces to a linear resistor and keeps the total conductance bounded away from zero. Built entirely from physical elements, the circuit is consistent with the conductance-based models of neurophysiology~\cite{hodgkin1952quantitative}.}
    \label{fig:model}
\end{figure}

The circuit model of Fig.~\ref{fig:model} with the mathematical formalism 
\begin{equation}{\label{eq:neuron_ODE}}
    C\dot{v} = -\sum_{i=1}^{p} g_i(v,w)\,\Bigl(v-E_i\Bigr) + u,
\end{equation}
is chosen as the representative model of neuromorphic systems. In this model, $C>0$ denotes the capacitance and is set to $1$ unless explicitly stated otherwise, and each term in the summation is a memristive branch of the circuit. For each one of the $p$ branches, $g_i(\cdot)$ denotes the conductance and is both nonlinear and dynamic. $w$ denotes auxiliary variables that the conductance might rely upon (such as gating variables for the Hodgkin--Huxley model \cite{hodgkin1952quantitative} or first-order lags of the voltage for mixed-feedback models \cite{ribar2019neuromodulation}).  $E_i$ is the reversal potential, which dictates whether the branch is excitatory or inhibitory, and $u$ is the injected current serving as input.

This model can be easily extended to \emph{networks} and incorporate synaptic connections, as
\begin{equation}{\label{eq:network_ODE}}
\begin{split}
\dot{v}_n = -\underbrace{\sum_{i=1}^{p} g_i(v_n,w_n)\,\Bigl(v_n-E_i\Bigr)}_{\text{Internal currents}} \\
-\underbrace{\sum_{\substack{j=1}}^{N} g_{jn}(v_j,w_j)\,\Bigl(v_n-E_{jn}\Bigr)}_{\text{Synaptic currents}}
+ u_n
\end{split}
\end{equation}
where $n$ highlights the index of the neuron of interest in a network of $N$ neurons and the synaptic current's conductances are modulated by the presynaptic voltages. Self-synapses are not permitted under this architecture, so $j\neq n$.

The adoption of this model is motivated by its several key advantages. First, its conductance-based modeling is compatible with the literature of neuroscience. This consistency prevents it from being a mere mathematical artifact. Second, the system is synthesized from fundamental circuit elements (capacitors, memristors, and batteries). Third, the model exhibits a rich collection of dynamical behaviors, including spiking and bursting. Finally, as a mathematical construct, this model comes with a structure that permits interpretations that intertwine physics and geometry \cite{forni2025gradient}.

\subsection{The Iterative Algorithm}

Equation \eqref{eq:network_ODE} presents a highly nonlinear and coupled dynamical system since the conductances $g_i$ and $g_{jn}$ are voltage-dependent. Suppose, however, that the conductances were evaluated not on the voltage, but on an estimate of it, namely $\hat{v}$ and $\hat{w}$. This results in conductance signals $g_i(t) \coloneqq g_i(\hat{v}_n,\hat{w}_n)$ and $g_{jn}(t) \coloneqq g_{jn}(\hat{v}_j,\hat{w}_j)$ that are voltage-independent and decouple \eqref{eq:network_ODE} into $N$ parallel scalar linear differential equations of the form
\begin{equation}
\dot{v}_n = -\sum_{i=1}^{p} g_i(t)\,\Bigl(v_n-E_i\Bigr) -\sum_{\substack{j=1}}^{N} g_{jn}(t)\,\Bigl(v_n-E_{jn}\Bigr)
+ u_n    
\end{equation}
that can be written more compactly as
\begin{equation}{\label{eq:simplified}}
    \dot{v}_n = -\eta (t)v_n + \xi  (t)   
\end{equation}
where $\eta (t) = (\sum_{i=1}^{p} g_i(t) + \sum_{\substack{j=1}}^{N} g_{jn}(t))$ and $\xi (t) = \sum_{i=1}^{p} g_i(t)\,E_i +\sum_{\substack{j=1}}^{N} g_{jn}(t)E_{jn}
+ u_n$. Defining $\Psi(t, s) = \exp\left(-\int_{s}^{t} \eta(\tau) \, d\tau\right)$, the closed form solution of \eqref{eq:simplified} is
\begin{equation}
    v_n(t) = \Psi(t, 0) v_{n,0} + \int_{0}^{t} \Psi(t, s) \xi(s) \, ds.
\end{equation}

By choosing the estimate $\hat{v}$ to be the voltage at the previous iteration $v^{k}$, this approach is turned into an iterative algorithm. The pseudocode for the proposed algorithm is provided below.
\begin{algorithm}[h]
\caption{Two-Step Iterative Solver}
\label{alg:twostep}
\begin{algorithmic}[1]
\State \textbf{Initialization:} $v^0 \in \mathcal{C}^N[0,T]$ with $v^0(0) = v_0$
\For{$k = 0,\dots,k_{\max}-1$}
    \State \textbf{Step 1:} $g_{i,n}^k \gets G_i\bigl(v_{n}^k\bigr), \qquad g_{jn}^k \gets G_{jn}\bigl(v_{j}^k\bigr)$
\For{$n = 1,\dots,N$}
    \State \textbf{Step 2:} Solve the LTV system of \eqref{eq:simplified}
    \[
    \dot{v}_n^{k+1} = -\sum_{i=1}^{p} g_{i,n}^k\,\Bigl(v_n^{k+1}-E_i\Bigr)  -\sum_{j=1}^{N} g_{jn}^k\,\Bigl(v_n^{k+1}-E_{jn}\Bigr) + u_n
    \]
\EndFor
    \If{$\lVert v^{k+1}-v^k\rVert_\infty \le \varepsilon$} 
        \State \textbf{break}
    \EndIf
\EndFor
\end{algorithmic}
\end{algorithm}

Algorithm \ref{alg:twostep} incur minimal computational cost. Specifically, the first step is a straightforward function evaluation, while the second is a linear time-varying ODE which has a closed-form integrating-factor solution. Section \ref{subsection3} establishes a guaranteed superlinear convergence rate. This algorithm can be physically interpreted as approximating memristive elements as frozen, time-varying linear conductors at each iteration, effectively reducing each step to the analysis of a linear RC circuit.

The next section reveals that this scheme is more than a heuristic: it can be derived via two distinct approaches that each bear its own interpretation, one being rooted in operator theory and the other in waveform relaxation.

\section{Interpreting the Algorithm}{\label{section3}}

\subsection{An Operator-Theoretic Perspective}{\label{subsection1}}

Simulating \eqref{eq:neuron_ODE} amounts to finding a voltage signal $v$ on $[0,T]$ that satisfies the differential equation together with its initial condition. Rather than viewing $v$ as a trajectory generated step by step from a state-space model, we treat the entire signal as a single unknown and the dynamics as an operator acting on it. The simulation problem then becomes a zero-finding problem in signal space:
\begin{equation}\label{eq:operator}
    \operatorname{W}(v) = 0 .
\end{equation}

For a single neuron, the operator $\operatorname{W}$ is
\begin{equation}\label{eq:W_def}
    \operatorname{W}(v) \coloneqq \partial v + \sum_{i=1}^{p} \operatorname{G}_i(v)\,\bigl(v - E_i\bigr) - u ,
\end{equation}
where $\partial$ denotes the differentiation operator, $\partial v \coloneqq \dot{v}$, and each $\operatorname{G}_i$ maps a voltage signal to the corresponding conductance signal $g_i(v,w)$. Because the auxiliary variables $w$ are themselves driven by $v$, the operator $\operatorname{G}_i$ absorbs their dynamics, so that the conductance depends on the voltage signal alone. All products in \eqref{eq:W_def} are taken pointwise in time. The operator acts on
\begin{equation*}
\begin{aligned}
    \operatorname{dom}(\operatorname{W}) \coloneqq \bigl\{ v \in \mathcal{L}_2^1[0,T] \;\big|\; &v \text{ absolutely continuous}, \\
    &\dot{v} \in \mathcal{L}_2^1[0,T],\ v(0) = v_0 \bigr\},
\end{aligned}
\end{equation*}
and maps into $\mathcal{L}_2^1[0,T]$. Encoding the initial condition in the domain ensures that the zeros of $\operatorname{W}$ are exactly the solutions of \eqref{eq:neuron_ODE}. The network \eqref{eq:network_ODE} is handled in the same way by stacking the voltages $v_1,\dots,v_N$ into a single signal in $\mathcal{L}_2^N[0,T]$ and adding the synaptic terms to $\operatorname{W}$.

Equation~\eqref{eq:operator} is nonlinear and, in general, admits no closed-form solution. Applying numerical methods to it directly is also not a tractable alternative to solve the problem in the signal space. The difficulty lies in the structure of $\operatorname{W}$, which places the linear dynamic operator $\partial$ alongside the nonlinear operators $\operatorname{G}_i$. Their sum inherits the difficulties of both, namely being simultaneously nonlinear and dynamic, and thus, no inexpensive method can resolve the combination.

Instead of treating $\operatorname{W}$ as a whole, one can decompose it into parts that are individually easier to handle and solve the original problem through a sequence of simpler subproblems. This divide-and-conquer strategy is the central idea of operator-splitting algorithms, which are well established in optimization theory~\cite{ryu2022algorithms}. The structure of $\operatorname{W}$ suggests a decomposition that follows the circuit itself. The first operator,
\begin{equation*}
    \operatorname{A} \coloneqq \partial ,
\end{equation*}
describes the current through the capacitor. It contains the lossless component of $\operatorname{W}$ and is linear. The second operator,
\begin{equation*}
    \operatorname{B}(v) \coloneqq \sum_{i=1}^{p} \operatorname{G}_i(v)\,\bigl(v - E_i\bigr) - u ,
\end{equation*}
collects the currents through the memristive branches together with the injected current. The system then takes the form
\begin{equation}\label{eq:splitting}
    \operatorname{A}v + \operatorname{B}(v) = 0 ,
\end{equation}
which is the standard setting addressed by splitting algorithms such as forward-backward splitting (FBS)~\cite{ryu2022algorithms}.

Applying the standard FBS algorithm to \eqref{eq:splitting} works for small circuits (although the classical monotonicity convergence results do not hold) but runs into numerical instabilities for large networks~\cite{shahhosseini2025operator}. Standard FBS measures distances in $\mathcal{L}_2$ with the same weight at every instant and for every neuron, whereas the conductances of a neuromorphic circuit vary sharply in time and across the network. Variable metric forward-backward splitting (VMFBS)~\cite{ryu2022algorithms} removes this rigidity by letting the geometry change from one iteration to the next. At iteration $k$, the Hilbert space is equipped with the inner product $\langle x,y\rangle_{\operatorname{M}^k} \coloneqq \langle \operatorname{M}^k x, y\rangle$, induced by a bounded, self-adjoint, strongly positive operator $\operatorname{M}^k$, and the VMFBS algorithm iterates
\begin{equation}\label{eq:VMFBS}
    v^{k+1} = \bigl(\operatorname{A} + \operatorname{M}^k\bigr)^{-1}\bigl(\operatorname{M}^k v^k - \operatorname{B}(v^k)\bigr).
\end{equation}

A good metric balances two competing goals. It should exploit the geometry of $\operatorname{B}$, so that the iterates converge quickly, while keeping the iterates inexpensive to compute. Variable metrics were first considered for neuromorphic circuits
in~\cite{shahhosseini2025variable}, which contains the seed of the present idea.
There, however, the metric was applied to only one of the two operators, so the resulting scheme was not, strictly speaking, a VMFBS method, and it came without convergence guarantees.

The metric is chosen to be the sum of the conductances at the current iterate. Writing $g_i^k \coloneqq \operatorname{G}_i(v^k)$, we define $\operatorname{M}^k$ as the multiplication operator
\begin{equation}\label{eq:metric}
    (\operatorname{M}^k x)(t) \coloneqq \sum_{i=1}^{p} g_i^k(t)\,x(t).
\end{equation}
It is assumed that the conductances are nonnegative and bounded, with their sum bounded below by some $\underline{g} > 0$, as is the case when the model includes a leak branch. Then $\operatorname{M}^k$ satisfies all the conditions that induce a valid metric. This metric has a direct physical meaning:
\begin{equation*}
    \|v\|_{M^k}^2
  = \int_0^T \sum_{i=1}^p g_i^k(t)\, v(t)^2 \,\mathrm{d}t
  = \bigl\langle v, \underbrace{M^k v}_{\text{current}} \bigr\rangle
\end{equation*}
is the energy dissipated in the memristive branches, frozen at their conductances at iteration $k$, when the voltage $v$ is applied across them. The metric thus measures a voltage error by the energy it would dissipate in the circuit, so errors at instants where the circuit is highly conductive weigh more than errors where it is nearly open.

Substituting \eqref{eq:metric} into \eqref{eq:VMFBS} makes the terms in $v^k$ cancel:
\begin{equation*}
    \operatorname{M}^k v^k - \operatorname{B}(v^k) = \sum_{i=1}^{p} g_i^k E_i + u .
\end{equation*}
Each iteration therefore solves
\begin{equation*}
    \Bigl(\partial + \sum_{i=1}^{p} g_i^k\Bigr) v^{k+1} = \sum_{i=1}^{p} g_i^k E_i + u ,
\end{equation*}
or, equivalently, the LTV system
\begin{equation}\label{eq:LTV_iteration}
    \dot{v}^{k+1} = -\sum_{i=1}^{p} g_i^k\,\bigl(v^{k+1} - E_i\bigr) + u , \qquad v^{k+1}(0) = v_0 .
\end{equation}
This is exactly Step~2 of Algorithm~\ref{alg:twostep}, with the conductances $g_i^k$ computed in Step~1. Algorithm~\ref{alg:twostep} is therefore VMFBS applied to \eqref{eq:splitting} with the metric \eqref{eq:metric}: each iteration freezes the memristors into time-varying linear resistors and solves the resulting RC circuit. The inverse in \eqref{eq:VMFBS} reduces to a linear time-varying ODE with a closed-form integrating-factor solution, and the energy-based metric adapts with the circuit at every iteration. Why this geometry yields fast convergence is made precise in the next subsections. Since $\operatorname{B}$ is not monotone, the convergence of Algorithm~\ref{alg:twostep} does not follow from classical monotone operator theory; it is established directly in Section~\ref{subsection3}.

The extension to a network of $N$ neurons is direct. We stack the voltages into $v = (v_1,\dots,v_N) \in \mathcal{L}_2^N[0,T]$, with $v_n(0) = v_{n,0}$, and define $\operatorname{A}$ and $\operatorname{B}$ componentwise:
\begin{equation*}
    (\operatorname{A}v)_n \coloneqq \partial v_n ,
\end{equation*}
\begin{equation*}
\begin{split}
    \bigl(\operatorname{B}(v)\bigr)_n \coloneqq{}& \sum_{i=1}^{p} \operatorname{G}_i(v_n)\,\bigl(v_n - E_i\bigr) \\
    &+ \sum_{j=1}^{N} \operatorname{G}_{jn}(v_j)\,\bigl(v_n - E_{jn}\bigr) - u_n .
\end{split}
\end{equation*}
The network \eqref{eq:network_ODE} then takes the same form \eqref{eq:splitting}, $\operatorname{A}v + \operatorname{B}(v) = 0$, now posed in $\mathcal{L}_2^N[0,T]$. The operator $\operatorname{A}$ is block diagonal, since it differentiates each voltage independently. All coupling between neurons resides in $\operatorname{B}$, through the synaptic conductances $\operatorname{G}_{jn}(v_j)$.

The metric \eqref{eq:metric} extends accordingly. Let
\begin{equation*}
    g_n^k \coloneqq \sum_{i=1}^{p} \operatorname{G}_i(v_n^k) + \sum_{j=1}^{N} \operatorname{G}_{jn}(v_j^k)
\end{equation*}
denote the total conductance seen by neuron $n$ at iteration $k$, and define
\begin{equation}\label{eq:metric_network}
    (\operatorname{M}^k x)_n(t) \coloneqq g_n^k(t)\, x_n(t), \qquad n = 1,\dots,N .
\end{equation}
Under the same assumptions on the conductances, $\operatorname{M}^k$ is bounded, self-adjoint, and strongly positive. Since $\operatorname{A} + \operatorname{M}^k$ is block diagonal, the inverse in \eqref{eq:VMFBS} decouples into $N$ scalar LTV ODEs,
\begin{equation*}
\begin{split}
    \dot{v}_n^{k+1} = {}& -\sum_{i=1}^{p} g_{i,n}^k \bigl(v_n^{k+1} - E_i\bigr) \\
    &- \sum_{j=1}^{N} g_{jn}^k \bigl(v_n^{k+1} - E_{jn}\bigr) + u_n ,
\end{split}
\end{equation*}
with $g_{i,n}^k \coloneqq \operatorname{G}_i(v_n^k)$ and $g_{jn}^k \coloneqq \operatorname{G}_{jn}(v_j^k)$. These are exactly the per-neuron updates in Step~2 of Algorithm~\ref{alg:twostep}, and they can be solved in parallel. Notably, the splitting involves two operators regardless of the number of neurons or their connectivity.

\subsection{A Waveform Relaxation Perspective}{\label{subsection2}}

The simulation of large-scale integrated circuits was a central challenge for the VLSI community in the late 1970s and early 1980s. \emph{Waveform relaxation} (WR)~\cite{white2012relaxation} was proposed to address it. Like Algorithm~\ref{alg:twostep}, it operates directly in signal space: its iterates are entire waveforms on $[0,T]$ rather than states at individual time steps. The idea is to decouple the system of ODEs describing the circuit and solve each equation separately. 

Concretely, consider a circuit described by
\begin{equation}\label{eq:WR_system}
    \dot{v}_n = f_n(v_1,\dots,v_N), \qquad v_n(0) = v_{n,0}, \qquad n = 1,\dots,N .
\end{equation}
The most basic WR scheme, Gauss--Jacobi WR, iterates
\begin{equation}\label{eq:jacobi_WR}
    \dot{v}_n^{k+1} = f_n\bigl(v_1^k,\dots,\, v_n^{k+1},\, \dots,v_N^k\bigr), \qquad v_n^{k+1}(0) = v_{n,0} .
\end{equation}
Each equation in \eqref{eq:jacobi_WR} is a scalar ODE in $v_n^{k+1}$ alone, driven by the frozen waveforms of the other variables. Because the $N$ equations are independent within an iteration, they can be solved in parallel.


The scalar equations in \eqref{eq:jacobi_WR} remain nonlinear, so each must itself be solved iteratively. A natural choice is Newton's method in signal space, which linearizes each equation around the current iterate. Combining WR with Newton's method in this way yields \emph{waveform relaxation Newton} (WRN) methods~\cite{white2012relaxation}. Their appeal lies in the fact that the inner Newton iteration need not be run to convergence. For the analogous one-step Jacobi--Newton method in finite dimensions, a single Newton step per outer iteration already attains the same asymptotic convergence rate as solving each subproblem exactly~\cite{ortega2000iterative}. Each WRN iteration therefore replaces a nonlinear subproblem with a linear time-varying one.

To make this precise, fix the waveforms $v_m^k$, $m \neq n$, from the previous iteration and define the residual operator of the $n$-th subproblem,
\begin{equation}\label{eq:WRN_ZFP}
    F_n(x) \coloneqq \dot{x} - f_n\bigl(v_1^k,\dots,v_{n-1}^k,\, x,\, v_{n+1}^k,\dots,v_N^k\bigr),
\end{equation}
acting on scalar signals $x$ with $x(0) = v_{n,0}$. Through the frozen waveforms, $F_n$ depends on the iteration index $k$. The Gauss--Jacobi WR update in \eqref{eq:jacobi_WR} is the zero of $F_n$. Rather than computing this zero exactly, WRN takes a single Newton step from the previous iterate $v_n^k$:
\begin{equation}\label{eq:WRN_Iterates}
    \mathrm{D}F_n(v_n^k)\bigl[v_n^{k+1} - v_n^k\bigr] = -F_n(v_n^k),
\end{equation}
where $\mathrm{D}F_n(v_n^k)$ is the Fr\'echet derivative of $F_n$ at $v_n^k$. Since both iterates satisfy the same initial condition, the increment $h \coloneqq v_n^{k+1} - v_n^k$ satisfies $h(0) = 0$. The derivative is
\begin{equation}\label{eq:Frechet_WRN}
    \mathrm{D}F_n(x)[h] = \dot{h} - \mathrm{D}_n f_n\bigl(v_1^k,\dots,x,\dots,v_N^k\bigr)[h],
\end{equation}
where $\mathrm{D}_n f_n$ denotes the Fr\'echet derivative of $f_n$ with respect to its $n$-th argument. When $f_n$ acts pointwise in time, $\mathrm{D}_n f_n[h]$ reduces to multiplication by $\partial f_n / \partial v_n$.

For the neuromorphic network \eqref{eq:network_ODE}, the synaptic conductances depend only on the frozen presynaptic voltages, so we write $g_{jn}^k \coloneqq \operatorname{G}_{jn}(v_j^k)$. The product rule then gives
\begin{equation}\label{eq:Jacobian_WRN}
    \begin{split}
    & \mathrm{D}F_n(x)[h] = \\
    & \qquad \dot{h} + \sum_{i=1}^{p} \Bigl( \operatorname{G}_i(x)\, h + \bigl(\mathrm{D}\operatorname{G}_i(x)[h]\bigr)\bigl(x - E_i\bigr) \Bigr) \\
    & \qquad {}+ \sum_{j=1}^{N} g_{jn}^k\, h .
    \end{split}
\end{equation}
The first term in parentheses is the conductance itself. The second is the sensitivity of the conductance to the voltage, which for gated conductances, requires propagating $h$ through the linearized gating dynamics. This second term is what makes full WRN expensive. Because the gates of conductance-based models have memory, $\mathrm{D}\operatorname{G}_i(x)[h](t)$ depends on the entire past of $h$ on $[0,t]$ rather than on $h(t)$ alone, and this can complicate the computations.

Quasi-Newton methods replace $\mathrm{D}F_n(v_n^k)$ in \eqref{eq:WRN_Iterates} with an approximation $\operatorname{P}_n^k$ and iterate
\begin{equation}\label{eq:WRQN}
    \operatorname{P}_n^k\bigl[v_n^{k+1} - v_n^k\bigr] = -F_n(v_n^k).
\end{equation}
Designing such methods is the art of balancing two goals: $\operatorname{P}_n^k$ should align with the geometry of the problem and simultaneously be cheap to compute. Dropping the sensitivity terms from \eqref{eq:Jacobian_WRN} gives
\begin{equation}\label{eq:QN_preconditioner}
    \operatorname{P}_n^k[h] = \dot{h} + \Bigl( \sum_{i=1}^{p} g_{i,n}^k + \sum_{j=1}^{N} g_{jn}^k \Bigr) h, \qquad g_{i,n}^k \coloneqq \operatorname{G}_i(v_n^k).
\end{equation}
Substituting \eqref{eq:QN_preconditioner} into \eqref{eq:WRQN} makes the terms in $v_n^k$ cancel and yields
\begin{equation*}
    \dot{v}_n^{k+1} = -\sum_{i=1}^{p} g_{i,n}^k\,\bigl(v_n^{k+1} - E_i\bigr) - \sum_{j=1}^{N} g_{jn}^k\,\bigl(v_n^{k+1} - E_{jn}\bigr) + u_n ,
\end{equation*}
which is exactly Step~2 of Algorithm~\ref{alg:twostep}. Algorithm~\ref{alg:twostep} is therefore a \emph{waveform relaxation quasi-Newton} (WRQN) method. For a single neuron, $\operatorname{P}^k = \partial + \operatorname{M}^k = \operatorname{A} + \operatorname{M}^k$. The variable metric \eqref{eq:metric} is thus precisely the quasi-Newton approximation of the derivative: it retains the curvature carried by the conductances and discards their sensitivity. Finally, the WRQN interpretation connects Algorithm~\ref{alg:twostep} to the extensive WR literature, which can be leveraged both for practical matters, such as choosing the discretization, and for theoretical ones, such as establishing convergence and its rate~\cite{white2012relaxation}.

\subsection{Convergence Analysis}{\label{subsection3}}

Algorithm~\ref{alg:twostep} offers an interpretable and principled approach towards simulating large-scale neuromorphic systems. This subsection proves that it converges to the solution of \eqref{eq:network_ODE}, uniformly on $[0,T]$ and superlinearly, and shows how the variable metric governs the rate.

With the inputs absorbed into the right-hand side, we write the network compactly as
\begin{equation}\label{eq:conv_system}
    \dot{v} = f(v), \qquad v(0) = v_0, \qquad f = (f_1,\dots,f_N).
\end{equation}
By \eqref{eq:network_ODE}, each term of $f_n$ is a conductance times a \emph{driving force} $v_n - E$:
\begin{equation}\label{eq:conductance_form}
\begin{aligned}
    f_n(v) = {}&-\sum_{i=1}^{p} G_i(v_n)\,(v_n - E_i) \\
    &- \sum_{j=1}^{N} G_{jn}(v_j)\,(v_n - E_{jn}) + u_n ,
\end{aligned}
\end{equation}
where $E$ ranges over the finite set $\mathcal{E}$ of reversal potentials, and the conductance operators $G_i$, $G_{jn}$ include the nonlinear filtering of the voltages. We denote by $g_n(v) \coloneqq \sum_{i} G_i(v_n) + \sum_{j} G_{jn}(v_j)$ the total conductance of neuron $n$, and $g_n^k \coloneqq g_n(v^k)$.

We consider preconditioners (consistent with Quasi-Newton methods) of the form $\operatorname{P}_n^k[h] = \dot{h} + c_n^k h$, where $c_n^k$ is a continuous scalar signal. Since $F_n(v_n^k) = \dot{v}_n^k - f_n(v^k)$, the update $\operatorname{P}_n^k\bigl[v_n^{k+1} - v_n^k\bigr] = -F_n(v_n^k)$ reads
\begin{equation*}
    \dot{v}_n^{k+1} - \dot{v}_n^k + c_n^k \bigl(v_n^{k+1} - v_n^k\bigr) = -\dot{v}_n^k + f_n(v^k).
\end{equation*}
Cancelling $\dot{v}_n^k$ gives the one-step scheme
\begin{equation}\label{eq:one_step}
    \dot{v}_n^{k+1} = f_n(v^k) - c_n^k \bigl(v_n^{k+1} - v_n^k\bigr), \qquad v_n^{k+1}(0) = v_{n,0},
\end{equation}
for $n = 1,\dots,N$. Algorithm~\ref{alg:twostep} corresponds to $c_n^k = g_n^k$, by \eqref{eq:WRQN}, and standard FBS with step size $\alpha$ to the constant $c_n^k = 1/\alpha$, by \eqref{eq:VMFBS} with $M^k = \alpha^{-1} I$.

\begin{assumption}\label{ass:lipschitz}
Problem \eqref{eq:conv_system} admits a solution $v^* \in \mathcal{C}^N[0,T]$, and there is a bounded interval $\mathcal{I} = [\underline{v}, \overline{v}]$ containing the values of $v^*$ and of all iterates, such that, for signals with values in $\mathcal{I}$:
\begin{enumerate}
    \item[(i)] every conductance operator $G \in \{G_i, G_{jn}\}$ is causal and Lipschitz, i.e., for all $t \in [0,T]$,
    \begin{equation*}
        \bigl\lvert G(x)(t) - G(y)(t) \bigr\rvert \le L_G \max_{\tau\in[0,t]} \bigl\lvert x(\tau) - y(\tau) \bigr\rvert ;
    \end{equation*}
    \item[(ii)] the conductances are nonnegative and bounded, and $g_n(v)(t) \ge \underline{g} > 0$ (e.g., due to the leak branch).
\end{enumerate}
\end{assumption}

On $\mathcal{I}$, the driving forces are bounded by
\begin{equation*}
    V_{\mathrm{d}} \coloneqq \max_{E \in \mathcal{E}}\, \max\bigl\{ \lvert \underline{v} - E \rvert,\, \lvert \overline{v} - E \rvert \bigr\},
\end{equation*}
and we define the \emph{sensitivity constant}
\begin{equation}\label{eq:sensitivity}
    L_s \coloneqq V_{\mathrm{d}} \max_{n} \Bigl( \sum_{i=1}^{p} L_{G_i} + \sum_{j=1}^{N} L_{G_{jn}} \Bigr),
\end{equation}
with $L_{G_{jn}} = 0$ if there is no synapse from $j$ to $n$. To interpret $L_s$, compare $f$ at an iterate $v^k$ and at the solution $v^*$, and let $e^k \coloneqq v^k - v^*$. For each branch $G(v)(v_n - E)$ of \eqref{eq:conductance_form}, where $G(v)$ stands for $G_i(v_n)$ or $G_{jn}(v_j)$, substituting $v_n^k - E = e_n^k + (v_n^* - E)$ gives
\begin{equation*}
\begin{aligned}
    G(v^k)&(v_n^k - E) - G(v^*)(v_n^* - E) \\
    &= G(v^k)\, e_n^k + \bigl[G(v^k) - G(v^*)\bigr](v_n^* - E).
\end{aligned}
\end{equation*}
Summing over the branches of neuron $n$ and using $g_n^k = \sum_i G_i(v_n^k) + \sum_j G_{jn}(v_j^k)$,
\begin{equation}\label{eq:f_split}
    f_n(v^k) - f_n(v^*) = -g_n^k\, e_n^k - r_n^k,
\end{equation}
where
\begin{equation*}
\begin{aligned}
    r_n^k \coloneqq {}&\sum_{i=1}^{p} \bigl[G_i(v_n^k) - G_i(v_n^*)\bigr] \bigl(v_n^* - E_i\bigr) \\
    &+ \sum_{j=1}^{N} \bigl[G_{jn}(v_j^k) - G_{jn}(v_j^*)\bigr] \bigl(v_n^* - E_{jn}\bigr).
\end{aligned}
\end{equation*}
By Assumption~\ref{ass:lipschitz}(i) and $\lvert v_n^* - E \rvert \le V_{\mathrm{d}}$, 
\begin{equation}\label{eq:r_bound}
    \lvert r_n^k(t) \rvert \le L_s \max_{\tau\in[0,t]} \lVert e^k(\tau) \rVert .
\end{equation}
A structure-informed Lipschitz constant of $f$ must therefore account for both the size of the conductances (corresponding to the first term of \eqref{eq:f_split}, containing $g_n^k$) and $L_s$ that corresponds to the second term. In Algorithm~\ref{alg:twostep}, the preconditioner cancels the first term exactly (Step~2 of the proof below), so only $L_s$ enters its rate.

\begin{theorem}\label{thm:convergence}
Let Assumption~\ref{ass:lipschitz} hold, let $v^0 \in \mathcal{C}^N[0,T]$ with $v^0(0) = v_0$, and let $c_n^k(t) \ge c_{\min} > 0$ with $\Delta \coloneqq \sup_{k,n,t} \lvert c_n^k(t) - g_n^k(t) \rvert < \infty$. Then, for every $\lambda \ge 0$, the errors $e^k$ of \eqref{eq:one_step} satisfy
\begin{equation}\label{eq:linear_rate}
\begin{aligned}
    \lVert e^{k+1} \rVert_\lambda &\le \gamma(\lambda)\, \lVert e^k \rVert_\lambda, \\
    \gamma(\lambda) &\coloneqq \bigl(\Delta + L_s\bigr)\, \frac{1 - e^{-(\lambda + c_{\min}) T}}{\lambda + c_{\min}},
\end{aligned}
\end{equation}
where $\lVert\cdot\rVert_{\lambda =0} = \lVert\cdot\rVert_\infty$. Consequently, $v^k \to v^*$ uniformly on $[0,T]$, and the convergence is R-superlinear:
\begin{equation}\label{eq:superlinear}
    \limsup_{k\to\infty} \lVert e^k \rVert_\infty^{1/k} = 0 .
\end{equation}
\end{theorem}

\begin{proof}

\emph{Step 1: Error dynamics.} Subtracting $\dot{v}_n^* = f_n(v^*)$ from \eqref{eq:one_step} and using $v_n^{k+1} - v_n^k = e_n^{k+1} - e_n^k$ gives
\begin{equation*}
    \dot{e}_n^{k+1} = f_n(v^k) - f_n(v^*) - c_n^k \bigl(e_n^{k+1} - e_n^k\bigr), \qquad e_n^{k+1}(0) = 0 .
\end{equation*}

\emph{Step 2: Conductance structure.} Substituting \eqref{eq:f_split} into Step 1,
\begin{equation}\label{eq:error_dynamics}
    \dot{e}_n^{k+1} = -c_n^k\, e_n^{k+1} + \bigl(c_n^k - g_n^k\bigr)\, e_n^k - r_n^k .
\end{equation}
The terms $-g_n^k e_n^k$ and $c_n^k e_n^k$ combine into the metric mismatch $(c_n^k - g_n^k)\, e_n^k$, which vanishes for Algorithm~\ref{alg:twostep}; only the conductance sensitivities $r_n^k$ remain.

\emph{Step 3: Forcing bound.} By definition of $\lVert\cdot\rVert_\lambda$, every $x \in \mathcal{C}^N[0,T]$ satisfies
\begin{equation}\label{eq:bielecki_pointwise}
    \lVert x(s) \rVert \le e^{\lambda s} \lVert x \rVert_\lambda \le e^{\lambda \tau} \lVert x \rVert_\lambda \qquad \text{for all } 0 \le s \le \tau .
\end{equation}
Applying \eqref{eq:bielecki_pointwise} to \eqref{eq:r_bound} and to $\bigl\lvert (c_n^k - g_n^k)\, e_n^k(\tau) \bigr\rvert \le \Delta\, \lVert e^k(\tau) \rVert$, the last two terms of \eqref{eq:error_dynamics} satisfy
\begin{equation}\label{eq:forcing_bound}
    \bigl\lvert (c_n^k - g_n^k)\, e_n^k(\tau) - r_n^k(\tau) \bigr\rvert \le (\Delta + L_s)\, e^{\lambda\tau}\, \lVert e^k \rVert_\lambda .
\end{equation}

\emph{Step 4: Variation of constants.} Equation \eqref{eq:error_dynamics} is a scalar LTV equation with zero initial condition, so
\begin{equation*}
\begin{aligned}
    e_n^{k+1}(t) &= \int_0^t \Phi_n^k(t,\tau) \bigl[(c_n^k - g_n^k)\, e_n^k - r_n^k\bigr](\tau)\, \mathrm{d}\tau, \\
    \Phi_n^k(t,\tau) &\coloneqq \exp\Bigl( -\int_\tau^t c_n^k(s)\, \mathrm{d}s \Bigr) \le e^{-c_{\min}(t-\tau)} .
\end{aligned}
\end{equation*}
With \eqref{eq:forcing_bound} and $\int_0^t e^{-c_{\min}(t-\tau)} e^{\lambda\tau}\, \mathrm{d}\tau = e^{\lambda t}\bigl(1 - e^{-(\lambda + c_{\min}) t}\bigr)/(\lambda + c_{\min})$,
\begin{equation*}
    \bigl\lvert e_n^{k+1}(t) \bigr\rvert \le (\Delta + L_s)\, \frac{1 - e^{-(\lambda + c_{\min}) t}}{\lambda + c_{\min}}\, e^{\lambda t}\, \lVert e^k \rVert_\lambda .
\end{equation*}

\emph{Step 5: Contraction of the error.} The right-hand side does not depend on $n$ and is nondecreasing in $t$. Maximizing over $n$, multiplying by $e^{-\lambda t}$, and taking the supremum over $t \in [0,T]$ gives $\lVert e^{k+1} \rVert_\lambda \le \gamma(\lambda) \lVert e^k \rVert_\lambda$. This concludes the proof of \eqref{eq:linear_rate}.

\emph{Step 6: Uniform and superlinear convergence.} The norm equivalence $e^{-\lambda T} \lVert x \rVert_\infty \le \lVert x \rVert_\lambda \le \lVert x \rVert_\infty$ and $k$ applications of \eqref{eq:linear_rate} give
\begin{equation}\label{eq:uniform_bound}
    \lVert e^k \rVert_\infty \le e^{\lambda T} \gamma(\lambda)^k\, \lVert e^0 \rVert_\infty .
\end{equation}
Since $\gamma(\lambda) \le (\Delta + L_s)/(\lambda + c_{\min}) \to 0$ as $\lambda \to \infty$, choosing $\lambda$ with $\gamma(\lambda) < 1$ in \eqref{eq:uniform_bound} shows that $\lVert e^k \rVert_\infty \to 0$. If $e^0 = 0$, \eqref{eq:superlinear} holds trivially. Otherwise, taking the $k$-th root of \eqref{eq:uniform_bound} gives $\limsup_{k\to\infty} \lVert e^k \rVert_\infty^{1/k} \le \gamma(\lambda)$ for every $\lambda \ge 0$, and $\inf_{\lambda \ge 0} \gamma(\lambda) = 0$ yields \eqref{eq:superlinear}.
\end{proof}

\begin{corollary}\label{cor:algorithm}
Let Assumption~\ref{ass:lipschitz} hold and $g^- \coloneqq \inf_{k,n,t} g_n^k(t) \ge \underline{g}$. Then Algorithm~\ref{alg:twostep} converges to the solution of \eqref{eq:network_ODE} uniformly on $[0,T]$ and R-superlinearly, with
\begin{equation}\label{eq:rate_alg}
    \gamma_{\mathrm{A}}(\lambda) = L_s\, \frac{1 - e^{-(\lambda + g^-) T}}{\lambda + g^-} .
\end{equation}
\end{corollary}

\begin{proof}
Algorithm~\ref{alg:twostep} is \eqref{eq:one_step} with $c_n^k = g_n^k$, so $c_{\min} = g^-$ and $\Delta = 0$. Theorem~\ref{thm:convergence} concludes the proof.
\end{proof}

\begin{remark}[Role of the metric]\label{rem:metric}
The decomposition \eqref{eq:f_split} splits the change in $f$ into two parts: the conductance term $-g_n^k e_n^k$ and the sensitivity term $r_n^k$. The variable metric of Algorithm~\ref{alg:twostep} cancels the conductance term exactly, so only the sensitivities drive the error and the mismatch $\Delta$ vanishes. A constant metric cannot track a conductance that changes by orders of magnitude during a spike, so standard FBS retains a mismatch of at least half the conductance range and obtains a looser bound. These are upper bounds, so they compare guarantees rather than observed rates; Fig.~\ref{fig:VMFBS} shows the same ordering in practice. A bound based only on a Lipschitz constant of $f$ would mix the two terms and could not reveal this difference. 

The factor $\gamma_{\mathrm{A}}$ in \eqref{eq:rate_alg} also has a physical reading. It is proportional to $L_s$, which measures how strongly the conductances react to voltage changes, and decreases with $g^-$, the smallest total conductance, i.e., the minimum dissipation of the circuit, which is lower bounded by the leak branch. A stronger leak dissipates energy faster, so errors fade faster and convergence accelerates.
\end{remark}

\begin{remark}
For Algorithm~\ref{alg:twostep}, the interval $\mathcal{I}$ of Assumption~\ref{ass:lipschitz} is explicit. If the total conductance is bounded below by $\underline{g} > 0$ for all voltages (e.g., by the leak), each update reads $\dot{v}_n^{k+1} = -g_n^k\bigl(v_n^{k+1} - \bar{E}_n^k\bigr) + u_n$, where $\bar{E}_n^k$ is a convex combination of the reversal potentials. Hence $\mathcal{I} = \bigl[E_{\min} - \lVert u\rVert_\infty/\underline{g},\, E_{\max} + \lVert u\rVert_\infty/\underline{g}\bigr]$ is forward invariant for every iterate and for the solution, provided that the initial condition and the initial guess take values in $\mathcal{I}$. For standard FBS, no such invariance is guaranteed, and the existence of $\mathcal{I}$ is part of the assumption.
\end{remark}

\begin{remark}
Theorem~\ref{thm:convergence} shows that the error $e^k = v^k - v^*$ contracts in $\lVert\cdot\rVert_\lambda$. It does not show that the iteration map $v^k \mapsto v^{k+1}$ is a contraction in the operator sense, which underlies the standard fixed-point analysis of splitting methods. That property would require the distance between the images of \emph{any} two signals to shrink, and Algorithm~\ref{alg:twostep} need not satisfy it. The analysis instead measures only the distance to the solution.
\end{remark}

\begin{remark}
Algorithm~\ref{alg:twostep} terminates on the increment in the supremum norm, not in the Bielecki norm used in Theorem~\ref{thm:convergence}. The two choices are consistent: the norms are equivalent on $\mathcal{C}^N[0,T]$, so convergence in one implies convergence in the other. The Bielecki norm is an analysis tool. Its weight $e^{-\lambda t}$ discounts late events in the simulation window and, for large $\lambda$, vanishes numerically. The supremum norm weighs all instants equally and stops the iteration once no part of the waveform changes by more than $\varepsilon$.
\end{remark}

\section{Example and Discussion}{\label{section4}}

\begin{figure}[h]
    \centering
    \includegraphics[width=0.95\linewidth]{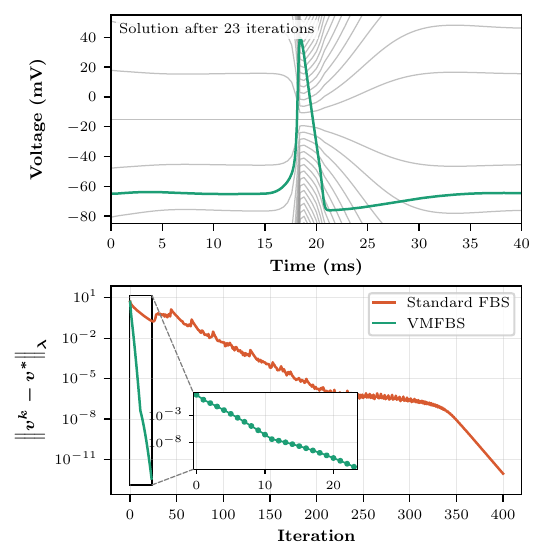}
    \caption{Hodgkin--Huxley neuron under a Gaussian input centered at 17.5 $\mathrm{ms}$. Top: Algorithm \ref{alg:twostep} converges after 23 iterations in the warped signal space provided by the geometry-induced variable metric (grey lines indicate equidistant coordinate lines). Bottom: error $\lVert v^k - v^*\rVert_\lambda$ with $\lambda = 1\,\mathrm{ms}^{-1}$ for standard FBS and Algorithm~\ref{alg:twostep} (VMFBS); the inset shows the first 20 iterations. Standard FBS has a weaker convergence guarantee (Remark~\ref{rem:metric}), and its error does not decrease monotonically. The variable metric weighs each instant by the total conductance, so instants where events occur receive more weight.}
    \label{fig:VMFBS}
\end{figure}

We illustrate Algorithm~\ref{alg:twostep} and its geometric interpretation on the Hodgkin--Huxley neuron~\cite{hodgkin1952quantitative}. The model, its parameters, and the solver are available in the accompanying code repository\footnote{\href{https://github.com/AmirShahhosseini/NeuromorphicNetworkSimulator}{https://github.com/AmirShahhosseini/NeuromorphicNetworkSimulator}}. Figure ~\ref{fig:VMFBS} illustrates the solution to the Hodgkin--Huxley neuron found using Algorithm \ref{alg:twostep} and how the space is warped with the variable metric, as discussed in Section \ref{subsection1}. It also shows the error of Algorithm~\ref{alg:twostep} decreasing monotonically in the Bielecki norm, consistent with Section \ref{subsection3}. Theorem~\ref{thm:convergence} holds for every $\lambda \ge 0$, contraction is guaranteed when $\gamma_{\mathrm{A}}(\lambda) < 1$, which holds when $\lambda > L_s - g^-$; this value is large, and over the simulation horizon the weight $e^{-\lambda t}$ would then vanish numerically. Fig.~\ref{fig:VMFBS} therefore uses $\lambda = 1\,\mathrm{ms}^{-1}$, and the observed monotone decrease is consistent with, but not guaranteed by, Corollary~\ref{cor:algorithm}. The solution found using Algorithm \ref{alg:twostep} agrees with the solution computed independently with a standard numerical integrator, as predicted by Corollary~\ref{cor:algorithm}.

\begin{figure}
    \centering
    \includegraphics[width=0.95\linewidth]{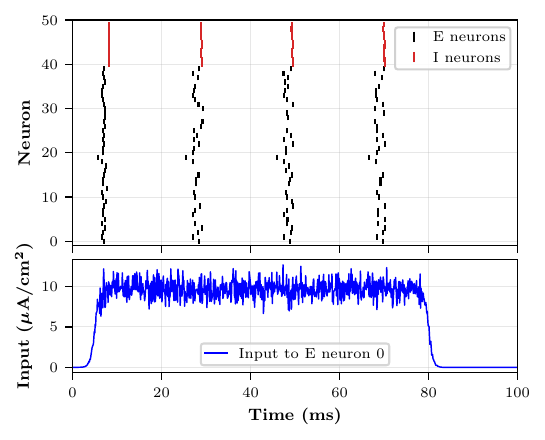}
    \caption{E--I network of 50 heterogeneous Hodgkin--Huxley neurons simulated with Algorithm~\ref{alg:twostep} ($\Delta t = 0.1\,\mathrm{ms}$). Top: raster plot of the 40 E neurons (black) and 10 I neurons (red), showing a PING-like rhythm. Bottom: external input to E neuron~0, a smooth pulse with frozen noise of standard deviation $10\%$ of its amplitude. Every neuron fires the same number of spikes as in a reference solution (DOP853, tolerance $10^{-10}$), with a median spike-time difference of $17.7\,\mu\mathrm{s}$.}
    \label{fig:network}
\end{figure}

To illustrate scalability, we simulate the excitatory--inhibitory (E--I) network of~\cite{Burger2023}, which generates pyramidal--interneuron gamma (PING) rhythms. Every E neuron excites every I neuron through a fast synapse, every I neuron inhibits every E neuron through a slow synapse, and only the E neurons receive an external input. Each synaptic conductance is scaled by $1/n_{\mathrm{pre}}$, the number of presynaptic neurons, so that the total synaptic drive of a neuron does not depend on the network size, permitting the seamless scaling of the network. The neurons are heterogeneous: their maximal conductances and input amplitudes are drawn with relative standard deviations of $10\%$, and the inputs carry frozen noise, in the spirit of~\cite{mainen1995reliability}. The complete model and its parameters are available in the code repository. Figure~\ref{fig:network} shows the raster plot of a network of 50 neurons. Every neuron fires the same number of spikes as in a reference solution computed with an adaptive eighth-order Runge--Kutta method (DOP853, tolerance $10^{-10}$), and the median spike-time difference is $17.7\,\mu\mathrm{s}$, consistent with the discretization error of the time step $0.1\,\mathrm{ms}$.

Algorithm~\ref{alg:twostep} is parallel by construction: within an iteration, the neurons interact only through the conductances frozen in Step~1, and Step~2 solves one independent scalar LTV equation per neuron. Assigning one GPU thread per neuron, a network of 5000 neurons converges in $9.28\,\mathrm{s}$ on the most basic GPU in Google Colab (T4 GPU). A systematic comparison with conventional simulators entails several intricacies and thus goes beyond the scope of this paper and is left for future work.

The iteration count is also consistent with Theorem~\ref{thm:convergence}. The sensitivity constant $L_s$ in~\eqref{eq:sensitivity} sums over the presynaptic inputs of each neuron. Since every synaptic conductance scales with $1/n_{\mathrm{pre}}$, this sum, and hence $L_s$, does not grow with the network size, and neither does $g^-$. The factor $\gamma_{\mathrm{A}}$ in~\eqref{eq:rate_alg}, and with it the number of iterations, should therefore be independent of $N$. Indeed, the networks of 50, 500, and 5000 neurons converge in 531, 558, and 567 iterations, respectively. These cold-start counts exceed the 23 iterations of Fig.~2 because the convergence is only asymptotically superlinear, and the transient lengthens with the simulation horizon (100\,ms vs. 40\,ms) and the number of spikes.

Networks raise an additional design choice: the order in which neurons are updated. Algorithm~\ref{alg:twostep} updates all neurons simultaneously from the previous iterate, which corresponds to the Gauss--Jacobi class of methods. This makes the updates fully parallel but ignores the network topology. Alternatively, neurons can be updated sequentially, with each update using the voltages of the already updated neurons, which corresponds to the Gauss--Seidel class of methods. A well-chosen ordering lets information propagate through the network within a single sweep. In a feedforward network, for instance, updating neurons in topological order means every neuron receives presynaptic inputs from the current iteration rather than the previous one. Such orderings can accelerate convergence substantially, as they incorporate the topology of the network within the solver \cite{white2012relaxation}.

\section{Conclusions and Future Work}{\label{section5}}

This paper introduced a geometry-aware, provably superlinear iterative solver for large-scale networks of conductance-based neurons. Its central message is that the very structure that makes these networks stiff, and thus hard to simulate, is also the key to making them tractable once built into the solver. Exploiting the physics of the circuit, rather than abstracting it away, is what lets the method scale.

Scalability, however, is not the only advantage of the approach. Because the solver operates in signal space, neuromorphic events can be captured at an adjustable level of detail without losing the behavior, a form of \emph{simulation at scale}. It enables the study of population-level behaviors that emerge directly from neuronal activity, bridging population coding and neuronal dynamics, a link still missing in neuroscience and neuromorphic engineering. Finally, methods of this class can incorporate prior knowledge into the solver, such as the shape of spikes or bursts, to further accelerate convergence, which is particularly valuable given the restricted repertoire of events in neuromorphic systems.

\bibliography{ref}
\bibliographystyle{ieeetr}

\end{document}